\documentclass[twocolumn, a4paper]{aastex62}

\usepackage{bm}
\usepackage{times}                    % Change the \texttt command to courier style
\usepackage{natbib}
 \bibpunct{(}{)}{;}{a}{}{,}
\usepackage{amssymb}                    % useful mathematical symbols
\usepackage{color}                       % For color text: \color command
\usepackage{url}                         % For breaking URLs easily trough lines
\usepackage{graphicx}                    % For eps figures, newer & more powerful

\def\beg{\begin{eqnarray}}
\def\ende{\end{eqnarray}}
\def\lsim{\lower.4ex\hbox{$\;\buildrel <\over{\scriptstyle\sim}\;$}}
\def\gsim{\lower.4ex\hbox{$\;\buildrel >\over{\scriptstyle\sim}\;$}}

\newcommand{\Rm}{\mbox{Rm}}

\renewcommand{\vec}[1]{\mbox{\boldmath $#1$}}

\def\Om{{\it \Omega}}

\def \Om  {{\it \Omega}}

\def\gsim{\lower.4ex\hbox{$\;\buildrel >\over{\scriptstyle\sim}\;$}} %$
\def\lsim{\lower.4ex\hbox{$\;\buildrel <\over{\scriptstyle\sim}\;$}} %$
\def\etaT{\eta_{\rm T}}
\def\um{u^{\rm m}}
\renewcommand{\vec}[1]{\mbox{\boldmath $#1$}}

\begin{document}

\title{Cycle period, differential rotation and meridional flow for early M dwarf stars\footnote{Based partly on data obtained with the STELLA robotic telescope in Tenerife, an AIP facility jointly operated by AIP and IAC}}
\shorttitle{Rotation and magnetic cycle times of early M dwarf stars}

\correspondingauthor{M.~K\"uker}
\email{mkueker@aip.de}

\author{M.~K\"uker}
\affiliation{Leibniz-Institut f\"ur Astrophysik Potsdam (AIP), An der Sternwarte 16, D-14482 Potsdam, Germany}

 \author{G.~R\"udiger}
 \affiliation{Leibniz-Institut f\"ur Astrophysik Potsdam (AIP), An der Sternwarte 16, D-14482 Potsdam, Germany}
 
 \author{K.~Ol\'ah}
\affiliation{Konkoly Observatory, Research Centre for Astronomy and Earth Sciences, Hungarian Academy of Sciences, Konkoly Thege 15-17, 1121, Budapest, Hungary }

\author{K.~G.~Strassmeier}
\affiliation{Leibniz-Institut f\"ur Astrophysik Potsdam (AIP), An der Sternwarte 16, D-14482 Potsdam, Germany}

\shortauthors{M.~K\"uker, G.R\"udiger, K. Olah, K.~G.~Strassmeier}

\date{Received; accepted}

\begin{abstract}
Recent observations suggest the existence of two characteristic cycle times for early-type M stars dependent on the rotation period. They are of order one year for the fast rotators ($P_{\rm rot}<1$~day) and of order 4 years for the slower rotators. Additionally, the equator-to-pole differences of the rotation rates with $\delta\Om$ up to 0.03~rad\,d$^{-1}$ are known from \emph{Kepler} data for the fast-rotating stars. These values are well-reproduced by the theory of large-scale flows in rotating convection zones on the basis of the $\Lambda$ effect. The resulting amplitudes $\um$ of the bottom value of the meridional circulation allows the calculation of the travel time from pole to equator at the base of the convection zone of early-type M stars. These travel times strongly increase with rotation period and they always exceed the observed cycle periods. Therefore, the operation of an advection-dominated dynamo in early M dwarfs, where the travel time {\em must} always be shorter than the cycle period, is not confirmed by our model nor the data.
%For the Sun, however,  cycle time and travel time coincide.
\end{abstract}


%\maketitle
%%%%%%%%%%%%%%%%%%%%%%%%%%%%%%%%%%%%%%%%%%%%%%%%%%%

\section{Introduction}

It is well-known \citep{R72} that an $\alpha\Om$ shell dynamo in its linear regime oscillates with a cycle time of
\beg
\tau_{\rm cyc} \simeq c_{\rm cyc} \frac{R_* D}{\etaT}
\label{I1}
\ende
with $R_*$ the stellar radius, $D$ the thickness of the convection zone, and $\etaT$ the turbulence-originated magnetic diffusivity of the convective flow.  Here the activity cycle is a diffusive phenomenon, that is only the volume of the convection zone and its diffusivity determine the oscillation frequency. The scaling factor $c_{\rm cyc}$ has been determined as $c_{\rm cyc}\simeq 0.26$ by means of shell dynamo models with constant shear \citep{RS72}. Adopting solar values for Eq.~({\ref{I1}}) leads to an Eddy diffusivity of $\etaT\simeq 10^{12}$~cm$^2$\,s$^{-1}$. With a more refined shell dynamo model \cite{K73} finds $\etaT\simeq 6\times 10^{11}$ cm$^2$\,s$^{-1}$  in close correspondence to the magnetic resistivity value derived from the decay of large activity regions at the solar surface \citep{SM90}. The analysis of the cross correlations $\langle \vec{u}\cdot\vec{b} \rangle$ taken from the solar surface as a function of the vertical large-scale surface field also provides  $\etaT\simeq 10^{12}$ cm$^2$\,s$^{-1}$ \citep{RK12}. Such low values for the Sun are not easy to understand because the canonical mixing-length estimate $u_{\rm rms}\ell_{\rm corr}$ exceeds these values by more than one order of magnitude ($u_{\rm rms}$ is the characteristic velocity and $\ell_{\rm corr}$ the characteristic correlation length of the turbulent convection). A measurement of $\etaT$ from the decay of starspots on a K0 giant suggested $6\times 10^{14}$~cm$^2$\,s$^{-1}$ \citep{KC15}.

% Ref:
% K�nstler, A., Carroll, T. A., \& Strassmeier, K. G. 2015, A\&A, 578, A101

Note that  Eq.~(\ref{I1}) does not show any dependence on the stellar rotation. Yet it is known from mean-field models that the dependence of the cycle time on the dynamo number reflects the strength of the back-reaction of the dynamo-generated field on the turbulence \citep{NW84,SS89}. An old and well-confirmed result is that the standard quadratic alpha-quenching concept leaves the cycle time uninfluenced. A more sophisticated 2D shell dynamo model with complete magnetic quenching of the turbulence-originated electromagnetic force leads to a weak and positive correlation of cycle and rotation periods in terms of $\tau_{\rm cyc} \propto P_{\rm rot}^{0.1}$ \citep{RK94}.
% Original Fig.1 and Fig. 3 are removed (they did not add anything new).

% New reference:
% Spada, F.; Demarque, P.; Kim, Y.-C.; Boyajian, T. S.; Brewer, J. M. 2017, ApJ, 838, 161
% Pojmanski, G., 1997, Acta Astronomica, 47, 467

The relation in Eq.~(\ref{I1}) allows predictions for the cycle length of cool main-sequence stars with outer convection zones. The total depth of the convection zone hardly varies with the spectral type. For $\etaT\simeq \mathrm{const}$ the cycle length for cool main-sequence dwarfs should basically vary only with $R_*$. The early M dwarfs with $R_*\simeq 0.4$~R$_\odot$ still have a shell-like outer convection zone while the M dwarfs cooler than  M3.5 are fully convective \cite[e.g.,][]{SD17}.
%(e.g., \citep{SD17})
A radius of $\approx 0.4$~R$_\odot$ leads to cycle times of about 4--5 years, provided that $\etaT$ remains unchanged. If $\etaT$ of cooler stars is reduced with respect to the solar value, as implicitly suggested by the stellar models of \cite{SD17}, then the cycle time from Eq.~(\ref{I1}) basically exceeds this limit.  Mixing-length estimates of $0.33 u_{\rm rms}\ell_{\rm corr}$ for M dwarfs provide about 30\%\ of the solar value, so that the dynamo cycle times of M dwarfs should slightly exceed the duration of the solar cycle.

\begin{figure}
\includegraphics[width=8.6cm]{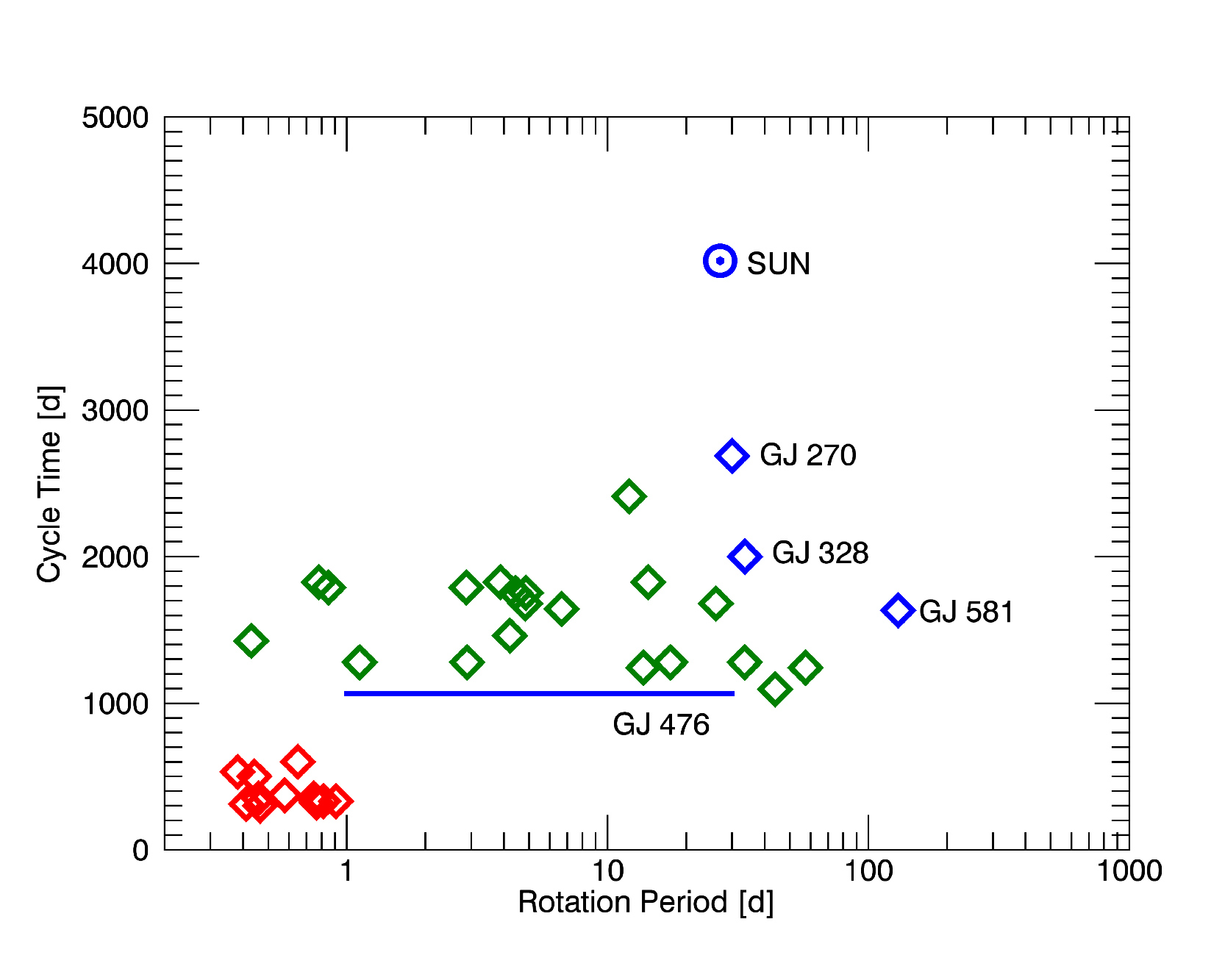}
\caption{Cycle times vs. rotation period (both in days) for early M stars (rhomb symbols). Among the group of rapid rotators with rotation periods less than a day \citep{VO14} are also the early M dwarfs EY~Dra and V405~And \citep{VK13}. The remaining data are from \cite{SR16,WS17,DL17}.  For  GJ476 the cycle time is known but not the rotation period (blue bar).
%The rotation period of GJ476 is still unknown.
%Note the solar-like cycle of the K2 dwarf HD\,219134 as a comparison.
The Schwabe cycle of the Sun is indicated as an asterisk marked SUN.}
\label{F1}
\end{figure}

% Fig. 1 does not show the K2 dwarf, add it! Also add the Sun 11 yr with an odot symbol.
%
Indeed, the cycle time of one star in Fig.~\ref{F1} slightly exceeds the 11-yr ($\approx$4000~d) limit but this turned out to be a K2 dwarf star with a rotation period of 22.8~d \citep{JE16}. Figure~\ref{F1} shows the observed cycle lengths for M dwarfs to be generally shorter than $\approx$3000~d. GJ270 (spectral type M2) shows the longest known cycle period of any M star with 2700~d (7.4\,yr), while the shortest cycle is seen in GJ476 (M4) with 1060~d (2.9\,yr) \citep{REb13}. GJ328 as an M0 dwarf with 0.69~M$_{\odot}$ shows an activity cycle of $\approx$2000~d, while a Jupiter-mass planet orbits it with a 4100-d period \citep{REa13}. Note that in the solar system the orbital period of Jupiter coincides with the length of the solar activity cycle, accidental or not. GJ581 possesses an entire planetary system with four or possibly six planets and shows an activity cycle with a length of 1600~d \citep{GS12,REb13}. The rotation period of the latter star is surprisingly long with $130\pm 2$~d \citep{RM14}.

\cite{GS12} give further examples of M dwarfs with cycle lengths between three and five years. It is of particular importance to know the relation of the cycle time with the rotation time for a homogeneous sample of stars. \cite{S12} used light curves of cool M dwarfs to find long-term cycle variations up to timescales of years. His sample is mixed by M stars with outer convection zones and also fully convective stars. There seems to be no systematic dependence of the cycle time on the rotation time even for subsamples with slow and fast rotation.

An independent study about the coupling of rotation and cycle times for fast rotating cool stars has been published by \cite{VO14}. Among the analyzed 39 \emph{Kepler} light curves of late-type stars with rotation periods shorter than one day, nine examples showed cyclic variations with periods between one and three years. Moreover, from  variations of the rotation periods, lower limits of the latitudinal differential rotation was derived with characteristic differences $\delta \Om\lsim 0.03$ rad\,d$^{-1}$.  Three of these targets (KIC04953358, KIC05791720, KIC10515986) have effective temperatures below 4000~K and can be considered early M dwarfs. Their minimum latitudinal shear $\delta \Om$ reaches from 0.008~rad\,d$^{-1}$ to 0.03~rad\,d$^{-1}$, which are values close to the theoretical results for ZAMS stars with $M=0.5$~M$_{\odot}$ by \cite{KR11}. The related meridional surface flow for fast rotation reaches amplitudes of 10~m\,s$^{-1}$.

In this paper, we have collected currently available data on cycle periods of M dwarfs from the literature and compare them with our numerical predictions. In Sect.~\ref{S2}, we reanalyze part of the All Sky Automated Survey (ASAS; \cite{P97}) data base for cycle periods, and present new STELLA photometry for three M dwarfs (\object{GJ270}, \object{GJ328}, \object{GJ476}) that had hitherto unknown rotational periods. In Sect.~\ref{S3}, we discuss the basic mechanisms of advection-dominated dynamos and present in Sect.~\ref{S4} new models for the large-scale flows in the outer convection zone of early M stars. Sect.~\ref{S5} summarizes our conclusions. The resulting surface values of the equator-to-pole differences of the angular velocity fit the observational data while the meridional flow amplitudes are basically too low to be responsible for the observed M-star cycle periods.

% --------------------------------------------- F1

\section{The observational data}\label{S2}

\subsection{Cycle periods}

From the literature we adopt only cycles from long-term changes of photometric time-series data with two constrains. Firstly, cycle periods are only accepted if the photometry was well sampled, that is only if there is no years-long interruption in the measurements which could introduce aliasing periods. Secondly, we require that the cycle period is not near the length of the data set itself, that is only if the changes were repeated or nearly repeated. A total of 25 targets were selected. For an earlier version of this approach, we refer to \cite{OS02}. All rotational and cycle periods used in this paper are listed in Table~\ref{table1}.

%For the fast-rotating \emph{Kepler} stars of early-M type we also list the values $\delta\Om$ found by \cite{VO14}.
%A characteristic value is 0.02 which is smaller than 0.06 resulting for the Sun (Fig. \ref{deltaom1}).

% New references:
% Kiraga, M. \&  Stepien, K. 2007, Acta Astron., 57, 149
% Kollath, Z. \& Ol\'ah, K. (2009), A&A, 501, 695
% Strassmeier, K. G. 2005, AN, 326, 269
% Ol\'ah, K. \& Strassmeier, K. G. 2002, AN, 323, 361
% Beer, J.; Tobias, S. M.; Weiss, N. O. 2018, MNRAS, 473, 1596

We re-examined the ASAS observations \citep{KS07} analyzed by \cite{S12} where many cycles were given for each star but the values themselves not published. We applied a time-frequency analysis on the same data sets with a bi-linear time-frequency transformation method like the one used by \cite{OK09}. For a description of this approach we refer to \cite{KO09}. Out of the available 31 targets, cycles for 19 targets could be verified and added to our sample. The results are given in Table \ref{table1} for stars of spectral type M3.5 or earlier. In most cases only a single cycle is found. For GJ729 and GJ897 two cycles were found but are harmonics where we can not decide which one is the correct one, while in case of GJ182, GJ2036A, GJ3367 and GJ803 the two detected cycle periods seem unrelated. The very fast rotating star GJ890 (=HK~Aqr) exhibits a single 3--4 yr cycle. Generally, the ASAS data do not allow constraining cycles of around one year due to the relatively low precision and low time resolution because the survey was designed for other purposes. Similar is the case for the fast rotating \emph{Kepler} stars, but then because of limited overall time coverage.

Since multiple cycles similar to the known solar cycles are also observed on stars \citep{OK09,OK16}, it is important to identify those stellar cycles which are true analogs of the dominating solar dynamo (Schwabe) cycle. The Sun seems to exhibit at least three cycle periods (but see \cite{BT18} for a full discussion); a mid-term cycle with 3--4~yr, the  Schwabe cycle with 9--14~yr, and the Gleissberg cycle with 90~yr or more. The Schwabe cycle is the by far dominating cycle and can be traced from the solar butterfly diagram showing the latitude migration of the sunspots in the course of a cycle. This is not directly doable for stars yet, but see the review by \cite{S05} for a summary of stellar cycle observations and tracers. 

Note  the many attempts of measuring stellar cycles from photometric time series. In   \cite{VO14} quasi-periodic changes of the rotational periods have been derived from the extremely precise \emph{Kepler} data sets,  based on the unresolved pattern of a solar-like butterfly diagram. Such  changes  on a number of fast rotating ($P_{\rm rot}<$~1 day) M dwarfs were found, which reflect the appearance of the spot's latitudes throughout an activity cycle. Therefore, we assume that the comparably short variability changes of the fast-rotating M dwarfs ($P_{\rm rot}<$~1day) in the sample of \cite{VO14} are most probably the true dynamo cycles of the stars.

Evidence of multiple cycles in fast-rotating M dwarfs are found for two well-observed stars from high precision ground-based data. V405~And and EY~Dra show cycle periods of less than a year \citep{VK13}, similar to the stars of the \emph{Kepler} field \citep{VO14}. However, the corresponding figure 1 from \cite{VO14} clearly shows, that these stars have additional, longer cyclic changes on timescales of, say, 4-5 years and/or even longer. The shorter cycles ($<$~1 year) of V405~And and EY~Dra are then likely the basic dynamo cycles similar to those of the \emph{Kepler}  stars. An example of a more massive star than an early M dwarf, but significantly lower mass than the Sun, is LQ~Hya (K2V). It has a rotational period of 1.6 days, whereas the Sun (G2V) has 27.24 days. The shorter cycle of LQ~Hya is about three years and corresponds to the solar 11-yr cycle given its dominance, the longer cycle of 7--12~yr may be an analog of the solar Gleissberg cycle (Ol\'ah \& Strassmeier 2002).

% --------------------------------------------- Table 1
\begin{table*}[tb]
\caption{Rotation periods, differential rotation  and cycle times of early-M dwarfs used in the present paper.}
\centering
\label{table1}
\begin{tabular}{llllll}
\hline\hline \noalign{\smallskip}
Star & Spectral type       & $P_{\rm rot}$ & $\delta\Om$& $P_{\rm cyc}$ & Cycle/rotation  reference  \\
     & or $T_{\rm eff}$ (K)&(days)         &(rad/day)  &(days)         &   \\
\noalign{\smallskip}\hline \noalign{\smallskip}
\object{GJ1054A} &	M0	&7.4&	&1278 &        present paper\\
\object{GJ1264A} &	M0.5&	6.66	&&1644 &        present paper\\
\object{HIP17695}	&M3&	3.88	&&1826   &      present paper\\
\object{GJ182}   &	M0.5	&4.43	&&1753, 1004 &  present paper\\
\object{GJ2036A} & 	M2&	2.98&	&1790,  913 &  present paper\\
\object{GJ205}	   &     M1.5	&33.5&	&1278  &       present paper\\
\object{GJ3331A} &	M1.3	&13.7	&&1242    &     present paper\\
\object{GJ3367}  &	M0&	12.1&	&2411, 1205  & present paper\\
\object{GJ358}	&        M2&	26.0	&&1680 &        present paper\\
\object{GJ431}   	&M3.5&	14.30&&	1826  &       present paper\\
\object{GJ494}   	&M0.5&	2.89&	&1278  &       present paper\\
\object{GJ618A}   &	M3&	57.4	&&1242 &        present paper\\
\object{GJ729}   &	M3.5&	2.87&&	1790 &   present paper\\
\object{GJ803}   	& M1	&4.86	&&1753, 1023 &  present paper\\
\object{GJ84}	   &     M2.5&	43.9&&	1096  &       present paper\\
\object{GJ841A}  	& M2.5&	1.12	&&1278  &       present paper\\
\object{GJ867A}  &	M1.5&4.22	&&1461  &       present paper\\
\object{GJ890} (HK Aqr)& 	M0&	0.43&&	1424  &       present paper \\
\object{GJ897A}  &	M2&	4.83&	&1680, 822&   present paper\\
\\
\hline
\\
\object{HD95735}	&  M2&	54.0&	&1424 & \cite{OK16}\\
GJ270&M2    &30 &&2687& \cite{REb13}, present paper\\
GJ476&M4    &\dots&&1066&\cite{REb13}\\
GJ328&M0    &33.6 &&2000&\cite{REa13}, present paper\\
GJ581&$>$M3 &130&&1633&\cite{GS12}\\
&&&&&\cite{REb13}\\
HIP1910  &       M0 &   1.75   & &1025&         \cite{DL16,DL17}\\
HIP23309&	M0&	8.74&	&1666 &         \cite{DL16,DL17}\\
HIP36349  &      M1 &   1.64 &  & 1069&    \cite{DL16,DL17}   \\
TWA2	&	M2&	4.85&	&1489&     \cite{DL16,DL17}     \\
TWA13A	&	M1&	5.44&	&1250 &     \cite{DL16,DL17}    \\
TYC5832-0666-1	&M0&	5.69&	&1695&   \cite{DL16,DL17}      \\
ProxCen &        M3.5&   83&  &	2557 &     \cite{WS17}  \\
GJ588 &	        M2.5 &   61.3&    &1899&   \cite{SR16}   \\
EY Dra &          M1.5 &   0.4587&  & 348 &   \cite{VK13}     \\
V405 And &        M0/M4&   0.4650&  & 300&   \cite{VK13}    \\
\\
\hline
\\
KIC 03541346   &4194  &   0.908154 &0.017 & 330     &\cite{VO14}\\
KIC 04819564  & 4125   &  0.380794 &0.0099 & 530   & \cite{VO14}\\
KIC 04953358  & 3843 &    0.649015 & 0.0048& 600      & \cite{VO14}\\
KIC 05791720   & 3533  &   0.765051 & 0.0098 &320     & \cite{VO14}\\
KIC 06675318  & 4206  &   0.577727 &0.0098 & 370     & \cite{VO14}\\
KIC 07592990 &  4004   &  0.442148  &0.0071 &500      & \cite{VO14}\\
KIC 08314902  & 4176 &    0.813534 & 0.0061& 330      & \cite{VO14}\\
KIC 10515986 &  3668  &   0.746207  &0.018& 350       & \cite{VO14}\\
KIC 11087527  & 4303   &  0.41096   &0.0076 &310      & \cite{VO14}\\
KIC 10063343 &3976 &0.3326 & 0.0057 & \dots &\cite{VO14}\\
\noalign{\smallskip}\hline
\end{tabular}
%\tablefoot{}
\end{table*}

\subsection{Rotation periods}

% Mallonn, M., Herrero, E., Juvan, I. G., et al. 2018, A\&A, in press

The new rotation periods in this paper also come from photometry. In particular, the rotation periods for the Gliese-Jahreiss (GJ) stars observed by ASAS are from the same data set as used for the cycle times, from simple Fourier analysis. The other periods are from the literature and referred to in Table~\ref{table1}.

Three GJ stars have a cycle period from the literature but not a rotation period \citep{REb13}. Therefore, these three stars (GJ270, GJ328, and GJ476) were put on the observing menu of one of our robotic STELLA telescopes in Tenerife for photometric monitoring. GJ270 and GJ328 were observed for a duration of four months, GJ476 for 40 nights, all of them in 2017/18. The STELLA-I telescope and its wide-field imaging photometer WiFSIP were employed. WiFSIP provides a field of view of 22'$\times$22' on a scale of 0.32\arcsec/pixel. The detector is a 4096$\times$4096 CCD with 15$\mu$m pixels. The observations were performed in blocks of five exposures in Johnson $V$ (150\,s exposure time) and five exposures in Cousins $I$ (60\,s). Data reduction and aperture photometry was done with the same ESO-MIDAS routines already used for similar monitoring programs of exoplanet host stars \citep{MN15,MH18}. The light curves versus Julian date are shown in the appendix in Sect.~\ref{A1}.

We applied the minimum string-length method described in \cite{D83} as well as a Lomb-Scargle periodogram to all three data sets (for two bandpasses each). None of the three targets exhibit variability significantly above the 1\% level in either bandpass which makes our periods tentative. GJ270 seems to be the best detection with a period of $\approx$30~d and an amplitude of 10\,mmag with a false alarm probability (fap) of 0.18\%. The Lomb-Scargle method gave additionally a period of 8.7~d which, however, is not seen from the minimum string length method and thus is rejected. A similar period of 33.6~d is obtained for GJ328 with a fap of 0.3\% and same amplitude. The Lomb-Scargle periodogram remained inconclusive for this target. The third star, GJ476, has no detection. Its largest minimizations are achieved with a very large fap of $\approx$7\% at periods of 11~d and 1.1~d but, more likely, the true period is longer than the 40 nights of observation. Again, its Lomb-Scargle periodogram did not show a significant peak.

%%%%%%%%%%%%%%%%%%%%%%%%%%%%%%%%
\section{Advection-dominated dynamos}\label{S3}

Meridional circulations influence the mean-field dynamo. This influence can be only a small modification if its characteristic time scale $\tau_{\rm m}$ basically exceeds the cycle time $\tau_{\rm cyc}$. Here, $u^{\rm m}$ represents the latitudinal flow close to the base of the convection zone. Positive $u^{\rm m}$ denotes a flow towards the equator. The cycle period becomes shorter for clockwise flow (in the first quadrant) but it becomes longer for counterclockwise flow \citep{RS72}  which is observed on the Sun. Fora critical  meridional circulations, when $\tau_{\rm m}\lsim \tau_{\rm cyc}$, the $\alpha\Om$ dynamo stops operation (in particular if the dynamo wave and the flux transport act in opposite directions).

% --------------------------------------------- F2
\begin{figure}
\includegraphics[width=8.6cm]{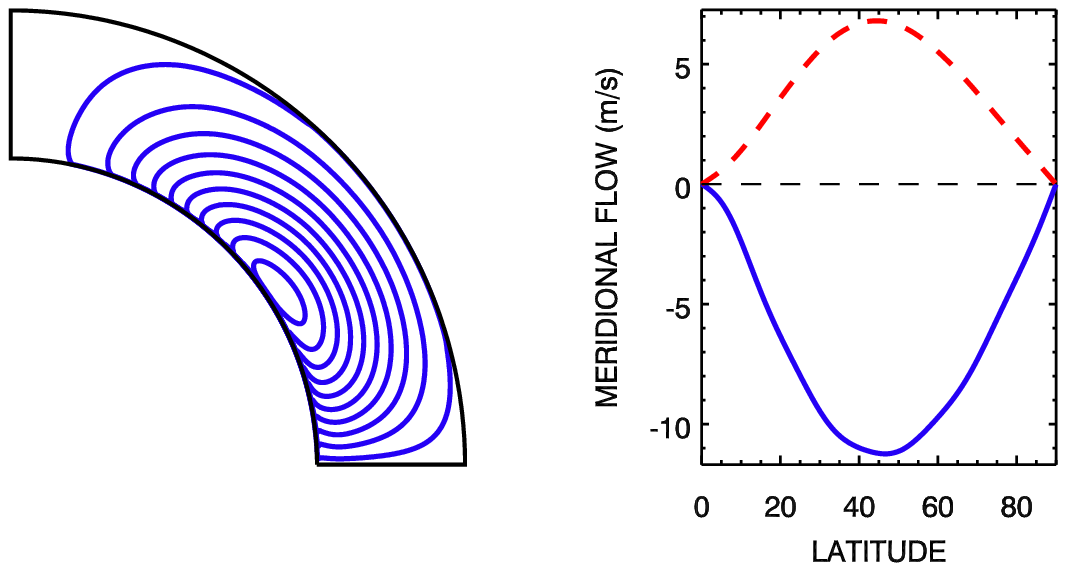}
\includegraphics[width=8.6cm]{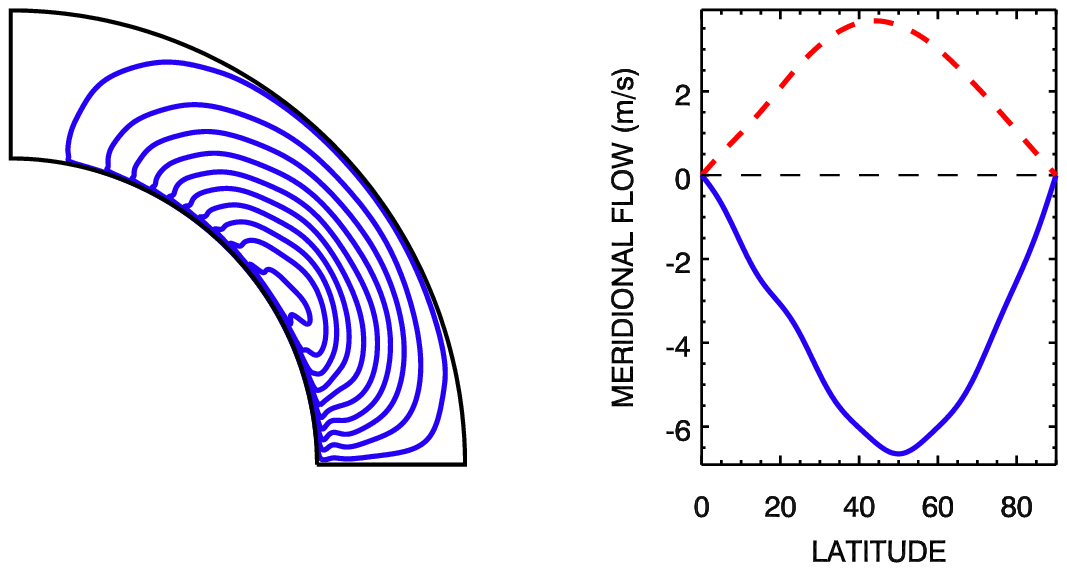}
\caption{Meridional flow system for two models for a star with a mass of $M=0.60$~M$_{\odot}$ and a rotation period of one day (top panels) and 10 days (bottom panels). We find poleward circulation at the top of the convection zone (blue full line) and equatorward flow at the base of the convection zone (red dashed line).}
\label{flow}
\end{figure}

For the solar cycle and with the estimate $\tau_{\rm m} \simeq R/\um$, we find $u^{\rm m}\simeq 2$~m\,s$^{-1}$ as a critical value above which the dynamo process is dominated by the flow and where in many models the cycle time varies as $1/\um$ \citep{DC99, KR01, BE02}.  Short cycles need fast flow  to be the result of an advection-dominated dynamo. Plausibly, the speed of the meridional circulation in convection zones grows with the rotation rate. Cycle time and rotation time are thus positively correlated in the mentioned dynamo models. The data for our early-M stars collected in Table~\ref{table1} and plotted in Fig. \ref{F1} do {\em not} contradict this condition. However, because standard $\alpha\Om$ dynamos can also show a (light) positive correlation of cycle time and rotation time \citep{RK94}, the shape of the function $\Om_{\rm cyc}(\Om)$ may not help  to find the nature of the operating dynamo.

A critical magnetic Reynolds number for advection-dominated dynamos is given by
\begin{equation}
 {\Rm}=\frac{\um D}{\etaT}\approx 10,
    \label{Rm}
\end{equation}
so that $\um=10$~m\,s$^{-1}$ is the minimal flow speed for $\etaT=10^{12}$~cm$^2$\,s$^{-1}$ and $D\approx 100\,000$~km.  It is the main question of this paper whether for the above-mentioned fast-rotating M stars the conditions for the operation of a flux-dominated dynamo are fulfilled. If we know the cycle times from observations it is only necessary to know the meridional flow amplitudes for the given rotation rates. Only if the cycles are longer than the turnover times of the meridional flow then an advection-dominated dynamo may operate. The ratio
\begin{equation}
 {\tau}=\frac{\tau_{\rm cyc}}{\tau_{\rm m}}
    \label{tau}
\end{equation}
exceeds unity for advection dominated dynamos if the travel time $\tau_{\rm m}$ is estimated by $\tau_{\rm m}\simeq \pi R_{\rm b}/u^{\rm m}$ with $R_{\rm b}$ the radius of the bottom of the convection zone. One finds $\tau\simeq 4$ for {\em all}  models of \cite{DC99,KR01,BE02}. For $\tau<1$ the operation of such a dynamo mechanism can be excluded.  It is now possible to calculate by means of the theory of differential rotation (see next section) the amplitudes of the meridional flow at the bottom of the convection zone. In particular the fast-rotating \emph{Kepler} stars in Fig.~\ref{F1} are interesting candidates for high enough flow speeds of the meridional circulation.
The travel time $\tau_{\rm m}$ at the bottom of the convection zone can be estimated with $\pi R_{\rm bot}/\um$ with $R_{\rm bot}$ as the radius of the base of the convection zone.
For M stars the travel time varies from 13 years for $\um\simeq 1$~m\,s$^{-1}$ to 1.3 years for $\um=10$~m\,s$^{-1}$. Therefore, one needs a rather precise theory of the meridional flow to calculate the true amplitude of $\um$.

\section{Differential rotation and meridional flow}\label{S4}

The theory of differential rotation in stellar convection zones provides simultaneous results for the rotation law and the meridional flow if the rotation rate is known \citep{RK13}. For the M dwarfs collected in Fig.~\ref{F1} the rotation rates are indeed known. Using spherical polar coordinates, one can reduce the Reynolds equation to a system of two partial differential equations. The azimuthal component of the Reynolds equation expresses the conservation of angular momentum,
\begin{equation} \label{omega}
   \nabla \cdot \vec{t} = 0,
\end{equation}
where $\vec{t}$ is the angular momentum flux vector with the components
\begin{equation}
t_i =  r \sin \theta (\rho r \sin \theta \Om {u}_i^{\rm m} + \rho Q_{i \phi}),
\end{equation}
with the angular velocity $\Om$, the meridional flow velocity $u^{\rm m}$, and the azimuthal components of the Reynolds stress, $Q_{\phi i}$.

The equation for the meridional flow is derived from the Reynolds equation by taking the azimuthal component of its curl,
\begin{equation}
 \label{curl}
  \left[ \nabla \times
	   \frac{1}{\rho}\nabla \cdot R \right ]_\phi
	   + r \sin \theta \frac{\partial \Om^2}{\partial z}
	  + \frac{1}{\rho^2}(\nabla \rho \times \nabla p)_\phi + \dots =0
\end{equation}
where $R_{ij}=-\rho Q_{ij}$ is the Reynolds stress and ${\partial}/{\partial z}$ is the derivative along the axis of rotation. The $\Lambda$ effect appears in two components of the Reynolds stress,
\begin{eqnarray}
 Q_{r \phi} &=& Q^{\rm visc}_{r \phi} + (V-H \cos^2\theta) \nu_{\rm T} \sin \theta \Om, \\
 Q_{\theta \phi} &=& Q^{\rm visc}_{\theta \phi} + H \nu_{\rm T} \sin^2\theta\cos \theta. \Om
 \label{lambda}
\end{eqnarray}
Here, $Q^{\rm visc}_{r \phi}$ and $Q^{\rm visc}_{\theta \phi}$ contain only first order derivatives of $\Om$ with respect to $r$ and $\theta$ and therefore vanish for uniform rotation. The coefficients $V$ and $H$ refer to the vertical and horizontal part of the $\Lambda$ effect.  The function $V$ is negative for slow rotation and becomes very small for fast rotation while $H$ is positive-definite and almost vanishes for slow rotation. Its existence seems to have been proven at the solar surface by SDO/HMI data \citep{RK14}. The Coriolis number $\Om^*=2 \tau_{\rm corr} \Om$  exceeds unity almost everywhere in the convection zone except in the outer layers. Close to $\Om^*=1$  it is $V\simeq H$ so that at the equator the radial angular momentum flux vanishes. The negativity of $V$ for slow rotation provides the negative shear in the solar super-granulation layer and it also allows anti-solar rotation for $\Om^*<1.$

The equations for the large-scale flows in convection zones have been solved for the Sun and for cool main-sequence stars \citep{KRK11,KR11} where the results for the pole-to-equator differences of $\Om$ and the surface values of the (poleward) meridional flow agree well with the observational data. At the bottom of the solar convection zone the resulting flow amplitude is about 6~m\,s$^{-1}$. The travel time of a gas parcel from pole to equator is thus about 6.5 years, shorter than the cycle time of 11 yrs. An advection-dominated dynamo for the Sun is thus possible but only if after (\ref{Rm}) the Eddy diffusivity does not exceed  $6\cdot 10^{11}$~cm$^2$\,s$^{-1}$ which is not easy to explain.

\begin{figure}
\includegraphics[width=8.6cm]{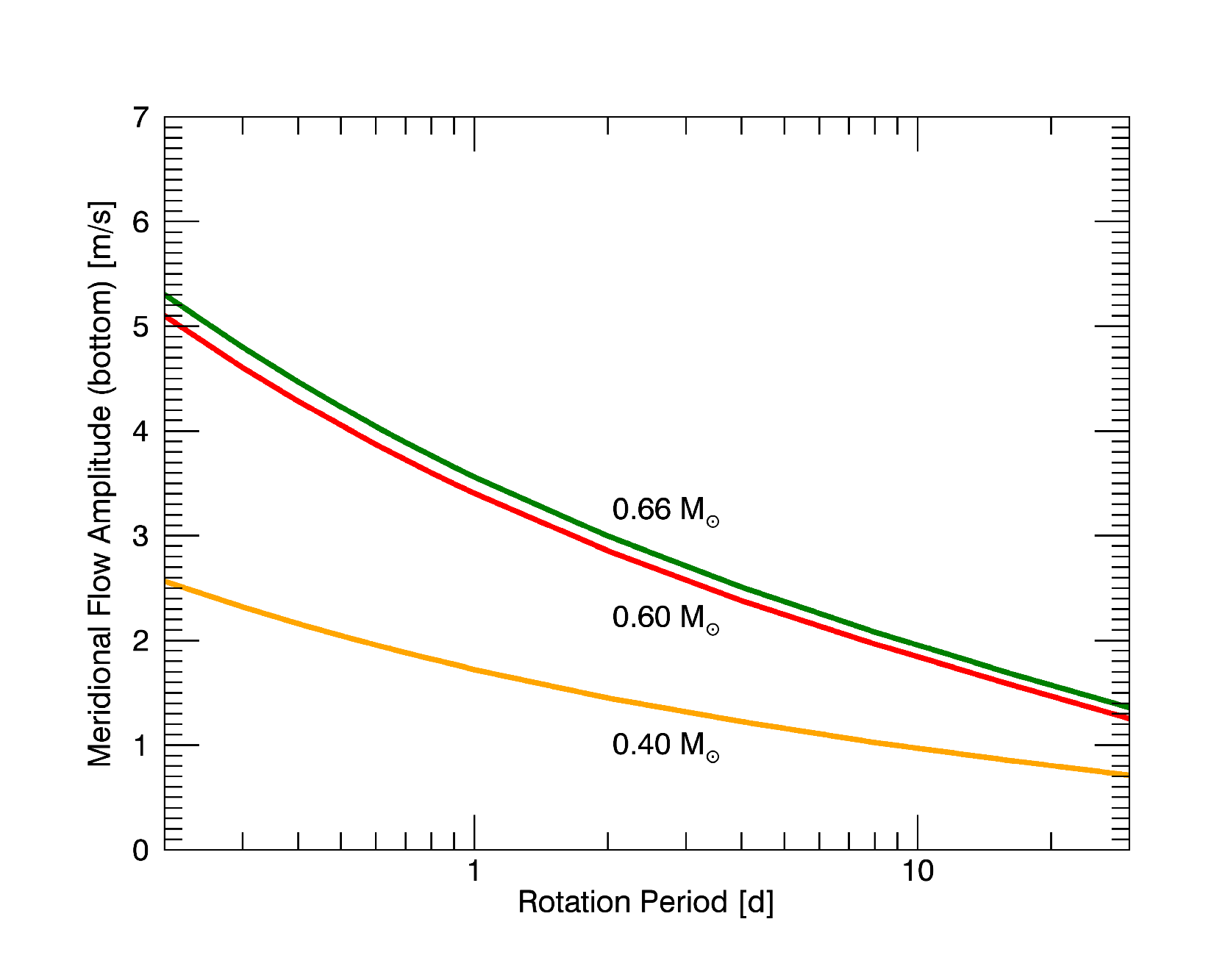}
\caption{Meridional flow amplitudes $\um$ averaged over the latitude at the bottom of the convection zones in m\,s$^{-1}$ for the  three different stellar models with masses of $M=0.40$~M$_{\odot}$ (yellow), $M=0.60$~M$_{\odot}$ (red) and $M=0.66$~M$_{\odot}$ (green). It refers to the rotation-law results in Fig.~\ref{deltaom2}.}
\label{botflow}
\end{figure}

For the meridional flows in M dwarfs the theoretical results are summarized in the Figs.~\ref{flow} and \ref{botflow}.  Figure~\ref{flow} shows the pattern of the meridional flow for rotation periods of one day (top panel) and 10 days (bottom panel). There is always one cell per hemisphere with the surface flow directed towards the poles and the return flow at the bottom of the convection zone. The amplitude of the return flow is between 4~m\,s$^{-1}$ for slow rotation and 7~m\,s$^{-1}$ for fast rotation.

% --------------------------------------------- F3

Figure~\ref{botflow} plots the circulation velocities at the bottom of the convection zones averaged over latitude for three solar metallicity ZAMS models which were computed with the MESA stellar evolution code \citep{PB11}. The masses were chosen to cover the spectral types of the stars in Tab.~\ref{table1}.
The meridional flows are fastest for rapid rotation and they grow with growing effective temperature. Above 4000~K, however, the  amplitudes $\um$ seem to saturate while for the M dwarfs below this temperature the $\um$ become basically smaller. For a rotation period of about one day a flow with $\um\simeq 3.5$~m\,s$^{-1}$ is obtained. Therefore, for such rapidly-rotating M dwarfs the travel time is 3.5 years or 1360~d, which is much longer than the observed cycle periods of about one year. At least, for the fast-rotating \emph{Kepler} stars given by \cite{VO14} the dynamo process cannot be of the advection-dominated type.

\begin{figure}
\includegraphics[width=8.6cm]{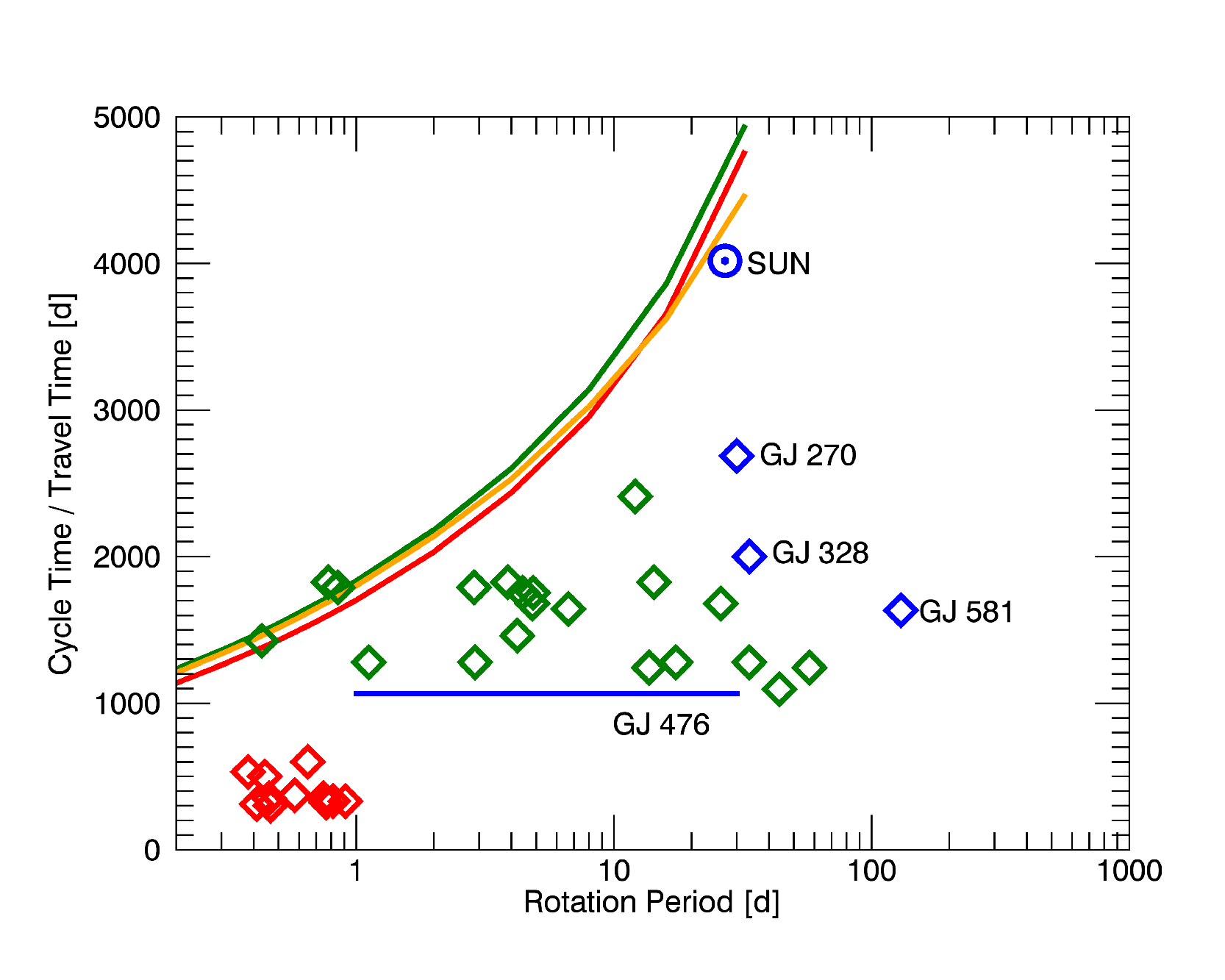}
\caption{Calculated travel times (in days, colored lines) and the observed cycle times for the early M stars from Fig.~\ref{F1} as a function of rotation period. The cycle times are always shorter than the calculated travel times, i.e. always $\tau\lsim 1$.  The data for the Sun
%and the K2-star HD\,219134 are
are shown for comparison. Colors as in Fig. \ref{botflow}. For GJ476 (blue bar) the cycle period is known but not the rotation period.}
\label{result}
\end{figure}

Figure~\ref{result} demonstrates that for all rotators the calculated travel times exceed the observed cycle times.  The cycles are too short to be originated by the meridional circulation. The circulation is thus only able to produce slight modifications in dependence on the rotation rate. Therefore, the existence of an advection-dominated dynamo is not confirmed by Fig.~\ref{result} for the sample of fast-rotating late-type stars.

% --------------------------------------------- F4
%  plot the Sun with an odot symbol in this plot

% you need to say something about the "three stellar models" first!!
As the stellar radii are smaller for cooler stars, the travel times for our three stellar models are almost identical (Fig.~\ref{result}).  %xxx this sentence makes no sense xxx. 
The travel times strongly increase with lower rotation, that is increase with rotation period. For all stars with masses 0.40--0.66~M$_{\odot}$ and with, say, solar rotation rate, the cycle time will be close to 11~years. None of the observed cycle times for early M stars lies above the single mixed-colored line, none of the observed cycle times exceeds the travel times from pole to equator for the early M stars so that the main condition for the existence of advection-dominated dynamos is nowhere fulfilled.

\begin{figure}
\includegraphics[width=8cm]{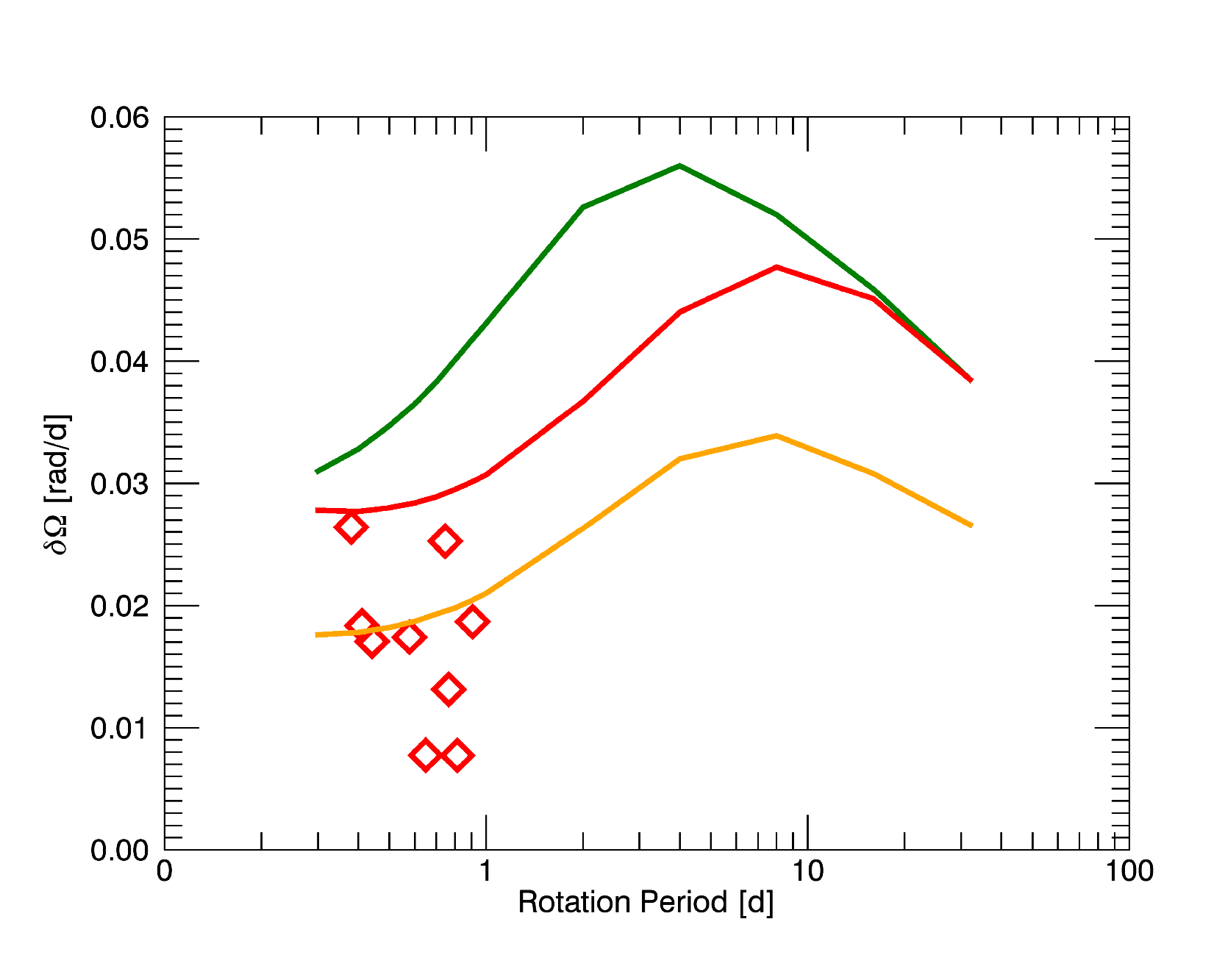}
\caption{Comparison of the shear values $\delta\Om$ from \cite{VO14} for fast-rotating M stars with our calculated values. The calculated values are shown as lines for stars of $M=0.66$~M$_{\odot}$ (green, top line), $M=0.60$~M$_{\odot}$ (red, middle line),  and $M=0.40$~M$_{\odot}$ (yellow, bottom line). }
\label{deltaom2}
\end{figure}

The evaluation of the meridional flow dynamics is only one side of the medal. The equation system simultaneously provides the rotation law of the convection zones. Only in case that also the theoretically resulting equator-to-pole differences of the stellar rotation rate complies with the observations, we can be relatively sure about the results for the meridional circulation inside the convection zone.  Figure~\ref{deltaom2} shows the influences of the stellar structure and rotation rate on the equator-to-pole difference of the angular velocity. The calculations are validated  for ZAMS models with masses of 0.66~M$_{\odot}$, 0.60~M$_{\odot}$, and 0.40~M$_{\odot}$. For $Z=0.02$ the models cover a temperature interval between 4325~K and 3581~K. The late K/early M stars are best described by $M=0.60$~M$_{\odot}$ and $T_{\rm eff}=4038$~K.

For all stellar  models the equator-to-pole differences of the surface rotation law grow with the rotation period if the rotation is not too slow.  The theoretical  shear grows for rotation periods of one day to ten days approximately by 50\%. For even slower rotation the rotational quenching of the $\Lambda$ effect leads to a reduction of the shear. The maximum of values of the differential rotation $\delta\Om$ appears for rotation periods slightly shorter than 10 days. For slower rotation the shear slowly decays by the rotational quenching of the $\Lambda$ effect. Note that there are even strong differences of the resulting $\delta\Om$ for the  two models with similar masses that, however, disappear for very fast and very slow rotation.

% --------------------------------------------- F5

One finds that the red line in Fig.~\ref{deltaom2} for a temperature of $T_{\rm eff}=4038$~K describes the behaviour of the early M stars best. The minimal value of the equator-to-pole difference is $\delta\Om\simeq 0.028$ for stars rotating slightly faster than one day. The result nearly perfectly fits the observations which in average yield {\em minimal} values. The  mean-field model of the hydrodynamic flows in the convection zone thus provides a correct interpretation of the differential rotation data  which may  support the reliability of  the results for the meridional circulation at the bottom of the convection zones.

%We are thus encouraged to discuss the meridional flow amplitudes at the bottom of the convection zones which belong to the solutions of the hydrodynamic equation system which for fast rotation correctly reflects the observed rotation laws of  the stellar surface.

\section{Conclusions}\label{S5}

We have shown that the observed activity cycles of the early-type M dwarfs are so short that they cannot be understood by an advection-dominated dynamo model. There is no single observation where the cycle period  remarkably exceeds the computed travel time from pole to equator. The hydrodynamical model on basis of the $\Lambda$ effect, which exists in rotating convection zones \cite[see][]{RK14}, provides a combination of differential rotation and meridional flow for a given stellar model. The computed differential rotation at the stellar surface complies with the observations for fast rotating \emph{Kepler} stars of spectral classes early M. It is thus allowed to use the meridional flow amplitudes, which are simultaneously provided by the model, as the characteristic amplitude for the flow pattern. As shown in Fig.~\ref{result} the travel times strongly depend on the stellar rotation period which, however, is not so obvious for the cycle times. The early-M stars do not fulfill the main condition of the advection dynamo concept that requires that the ratio in Eq.~(\ref{tau}) exceeds unity.

\acknowledgements
K.O. thanks for the support from the MTA CSFK Discretional Fund.

\bibliography{cycles}

\appendix

\section{STELLA photometry of GJ270, GJ328, and GJ476}\label{A1}

\begin{figure}[h]
\includegraphics[width=8.6cm,angle=0]{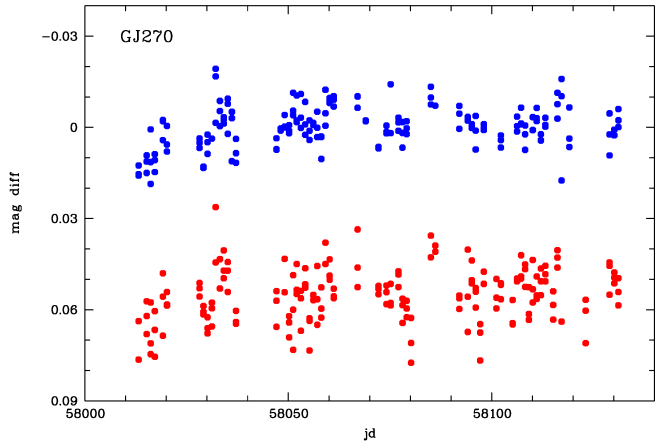}
\caption{Light curve of GJ270. Time is in Julian dates plus 2,400,000. Shown is the differential magnitudes for the $V$ band (blue dots, upper band) and the $I_C$ bad (red dots, lower band). }
\label{A-F1}
\end{figure}

\begin{figure}[h]
\includegraphics[width=8.6cm,angle=0]{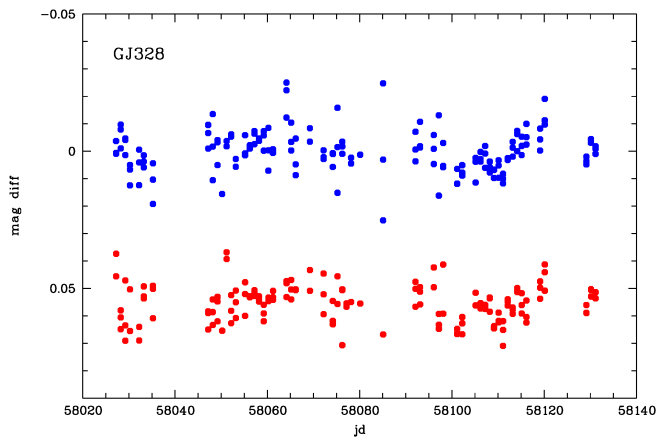}
\caption{Light curve of GJ328. Otherwise as in Fig.~\ref{A-F1}.}
\label{A-F2}
\end{figure}

\begin{figure}[h]
\includegraphics[width=8.6cm,angle=0]{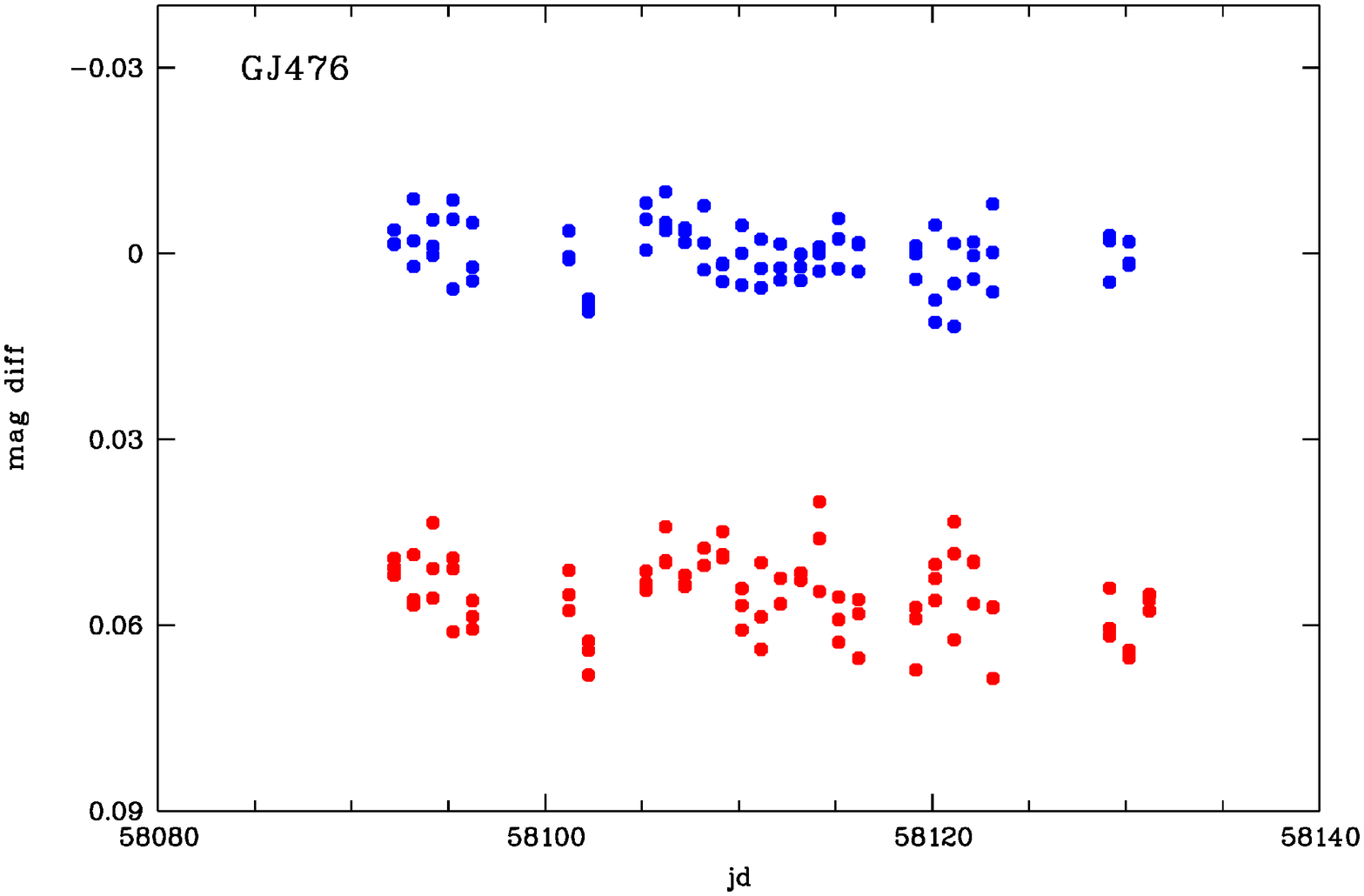}
\caption{Light curve of GJ476. Otherwise as in Fig.~\ref{A-F1}.}
\label{A-F3}
\end{figure}

\end{document}